\documentclass[useAMS,usenatbib,usegraphicx]{mn2e}
\usepackage{amssymb}

\title[Detecting lensed pop III galaxies with HST and JWST]{Detecting gravitationally lensed population III galaxies with HST and JWST}
\author[Zackrisson et al.]{Erik Zackrisson$^{1}$\thanks{E-mail: ez@astro.su.se}, Adi Zitrin$^{2}$, Michele Trenti$^{3}$, Claes-Erik Rydberg$^{1}$, Lucia Guaita$^{1}$,\newauthor Daniel Schaerer$^{4,5}$, Tom Broadhurst$^{6,7}$, G\"oran \"Ostlin$^{1}$, Tina Str\"om$^{1}$\vspace{0.25cm}\\  
$^{1}$Department of Astronomy, Stockholm University, Oscar Klein Center, AlbaNova, Stockholm SE-106 91, Sweden\\
$^{2}$Institut f\"{u}r Theoretische Astrophysik, ZAH, Albert-Ueberle-Stra\ss e 2, 69120 Heidelberg, Germany\\
$^{3}$Institute of Astronomy, University of Cambridge, Madingley Road, Cambridge CB3 0HA, UK\\
$^{4}$Geneva Observatory, University of Geneva, 51 Chemin des Maillettes, CH-1290 Versoix, Switzerland\\
$^{5}$CNRS, IRAP, 14 Avenue E. Belin, F-31400 Toulouse, France\\
$^{6}$Department of Theoretical Physics, University of Basque Country\\
$^{7}$IKERBASQUE, Basque Foundation for Science}

\begin{document}

\date{Accepted ... Received ...; in original form ...}

\pagerange{\pageref{firstpage}--\pageref{lastpage}} \pubyear{2010}

\maketitle

\label{firstpage}

\begin{abstract}
Small galaxies consisting entirely of population III (pop III) stars  may form at high redshifts, and could constitute one of the best probes of such stars. Here, we explore the prospects of detecting gravitationally lensed pop III galaxies behind the galaxy cluster J0717.5+3745 (J0717) with both the {\it Hubble Space Telescope} ({\it HST}) and the upcoming {\it James Webb Space Telescope} ({\it JWST}). By projecting simulated catalogs of pop III galaxies at $z\approx 7$--15 through the J0717 magnification maps, we estimate the lensed number counts as a function of flux detection threshold. We find that the ongoing {\it HST} survey {\it CLASH}, targeting a total of 25 galaxy clusters including J0717, potentially could detect a small number of pop III galaxies if $\sim 1\%$ of the baryons in these systems have been converted into pop III stars. Using {\it JWST} exposures of J0717, this limit can be pushed to $\sim 0.1\%$ of the baryons. Ultra-deep {\it JWST} observations of unlensed fields are predicted to do somewhat worse, but will be able to probe pop III galaxies with luminosities intermediate between those detectable in {\it HST/CLASH} and in {\it JWST} observations of J0717. We also explain how current measurements of the galaxy luminosity function at $z=7$--10 can be used to constrain pop III galaxy models with very high star formation efficiencies ($\sim 10\%$ of the baryons converted into pop III stars).
\end{abstract}


\section{Introduction}
\label{intro}
The first, metal-free (population III) stars are predicted to form in isolation or in small numbers within dark matter minihalos of mass $10^{5-6}\ M_\odot$ at $z\approx 10-50$ \citep[e.g.][]{Tegmark et al.,Yoshida et al.,Trenti & Stiavelli 09,Stacy et al.}. Some of these population III stars (hereafter pop III) likely explode as supernovae and enrich the surrounding medium with heavy elements, initiating the transition to the population II/I star formation modes known from the low-redshift Universe. 

As more massive dark matter halos ($10^{7-8}\ M_\odot$) start to assemble, small galaxies capable of sustaining prolonged star formation probably emerge at $z\approx 10$--15 \citep[e.g.][]{Greif et al. 08,Johnson et al. 09}. Since detailed simulations suggest that most of these galaxies should form in high-density regions that were pre-enriched by pop III stars in minihalos, these galaxies are not expected to be completely metal-free \citep[e.g.][]{Greif et al. 10} and are most likely dominated by chemically enriched stars. True pop III galaxies may, however, still form in low-density environments where Lyman-Werner feedback has prevented prior star formation in minihalos, and into which supernova ejecta from adjacent regions has not yet migrated \citep[e.g.][]{Scannapieco et al.,Tornatore et al.,Trenti et al. 09,Stiavelli & Trenti 10}.

Simulations of the collapse and fragmentation of metal-free gas generically predict that pop III stars should form with high characteristic masses \citep[for a review, see][]{Bromm & Larson}, but the exact initial mass function of these objects remains poorly constrained. About a decade ago, many studies favoured pop III masses on the order of $\sim 100\ M_\odot$ \citep[e.g.][]{Bromm et al.,Abel et al.}, but the most recent simulations hint at masses closer to $\sim 10\ M_\odot$ and an initial mass function that may even extend below $\sim 1 \ M_\odot$ \citep[e.g.][]{Stacy et al.,Clark et al.,Greif et al. 11a,Greif et al. 11b,Hosokawa et al.}. Since isolated pop III stars are likely to be outside the reach of even the James Webb Space Telescope\footnote{http://www.jwst.nasa.gov/} \citep[e.g.][]{Gardner et al.,Greif et al. 09,Rydberg et al.}, at least before they go supernovae \citep{Weinmann & Lilly,Whalen & Fryer}, pop III galaxies may offer one of the best routes to observationally pinning down the properties of pop III stars. 

While \citet{Inoue} and \citet{Zackrisson et al. b} have argued that pop III galaxy candidates can be singled out from {\it James Webb Space Telescope} ({\it JWST}) multiband surveys because of their unusual spectra, the intrinsic luminosities of these objects may be too low for detection in even the longest {\it JWST} exposures \citep{Johnson et al. 09,Stiavelli & Trenti 10}. Pop III galaxies at $z\approx 6$--10 can, in principle, also be identified in {\it HST} imaging surveys \citep{Inoue,Zackrisson et al. c}, although this places even harsher demands on their luminosities. Here, we will explore to what extent gravitational lensing may brighten the prospects for detection.

Lensing by foreground, low-redshift galaxy clusters can boost the fluxes of high-redshift objects by factors of $\mu\sim 10$--100 \citep[e.g.][]{Bradley et al. a,Zheng et al.,Bradac et al.,Maizy et al.,Zitrin et al. a,Zitrin et al. b,Hall et al., Bradley et al. b,Zitrin et al. c}, but at the expense of zooming in on a high-redshift volume that is smaller by the same factor. Because of these competing effects, surveys targeting lensed fields are not equally suitable for all source populations \citep[e.g.][]{Maizy et al.,Bouwens et al. 09}. 

In this paper we investigate whether high-redshift ($z\gtrsim 7$) pop III galaxies behind the $z=0.546$ galaxy cluster MACS J0717.5+3745 (hereafter J0717) may be detected with either the {\it HST} or {\it JWST}. This cluster has the largest angular Einstein radius detected so far \citep[$\simeq55\arcsec$ for a source at $z_{s}\sim2.5$;][]{Zitrin et al. a}. Due to the complex merging nature of this cluster, it has a relatively shallow surface mass-density profile which boosts the projected area over which high magnifications are attained \citep{Zitrin et al. a,Zitrin et al. b}. Both of these properties combine to make this object one of the best lenses currently known (quite possibly {\it the} best) for the study of faint objects at very high redshifts. Previously, we have demonstrated that $z\sim 10$ dark stars could in principle be detectable in a {\it JWST} imaging survey targeting this object \citep{Zackrisson et al. a}.

Our simulations are presented in Sect.~\ref{simulation}.  In Sect.~\ref{Palantir}, we explore the prospects of detecting lensed pop III galaxies at $z\gtrsim 7$ in a hypothetical {\it JWST} survey of J0717. The merits of this approach are also compared to those of ultradeep {\it JWST} observations of unlensed fields. Sect.~\ref{HST_section} contains the corresponding analysis for the {\it HST/CLASH} survey \citep{Postman et al.}, which targets a total of 25 galaxy clusters, including J0717. Constraints on the properties of pop III galaxies based on existing observations of the galaxy luminosity function at $z=7$--10 are presented in Sect.~\ref{LF}. A number of uncertainties in our analysis are discussed in Sect.~\ref{discussion}. 

\section{Mock samples of population III galaxies}
\label{simulation}

\begin{figure}
\includegraphics[width=84mm]{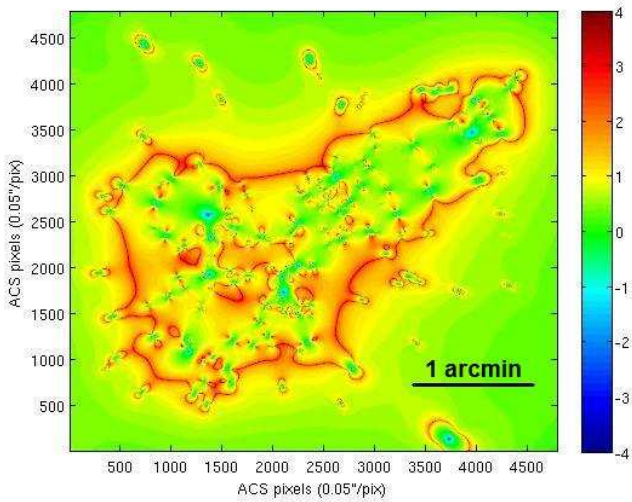}
\caption{$4\times 4$ arcmin magnification map for the largest lens currently known, the J0717 galaxy cluster \citep{Zitrin et al. a}. The color bar on the right shows the $\log_{10}(\mu)$ magnification values plotted in the figure, assuming a source redshift of $z\simeq10$.
\label{J0717}}
\end{figure}

The galaxy cluster J0717 at $z=0.546$ (see Fig.~\ref{J0717}) is a very unusual object, with an unrelaxed morphology, a correspondingly shallow density profile and the largest Einstein radius so far detected \citep{Zitrin et al. a}. The area over which at $z>6$ background objects attain high magnifications $\mu$ is exceptionally large: $\approx 3.5$ arcmin$^2$ for $\mu>10$. This potentially makes it the best lensing cluster currently known for studying faint objects at very high redshifts. J0717 is one of the targets of the ongoing {\it HST/CLASH}\footnote{http://www.stsci.edu/$\sim$postman/CLASH/Home.html} survey, covering 25 galaxy clusters in 16 filters, and also an exciting target for upcoming studies with {\it JWST}.

To explore the prospects of searching for pop III galaxies in a highly magnified field like J0717, and to weight the merits of this approach against observations of unlensed fields, we generate mock pop III galaxy catalogs for both of these situations. The computational machinery for this involves the following steps:

\begin{enumerate}
\item For the lensed samples, simulated pop III halo catalogs (see Sect.~\ref{halos}) are projected through the cluster magnification maps (Sect.~\ref{lensing}). This produces the positions and magnifications of all potential pop III galaxy images within the J0717 field. For the unlensed samples, the pop III halo catalogs are simply left as they are.
\item Each distinct halo in these catalogs is assigned a random stellar population age (Sect.~\ref{MC}).
\item Using the spectral synthesis model Yggdrasil \citep{Zackrisson et al. b}, we then derive the {\it HST} and {\it JWST} fluxes of each pop III galaxy image within these mock samples, as a function of various model parameters (see Sect.~\ref{SEM}). The fluxes in the lensed samples are adjusted by a factor equal to the magnification of each separate image. Due to the image splitting associated with strong lensing, a single pop III galaxy can appear at several locations within the J0717 field, each with a different magnification.
\end{enumerate}

\subsection{Population III halos}
\label{halos}
We construct Population III halo catalogs based on the numerical
simulations of \citet{Trenti et al. 09}. To summarize, we run a
high-resolution cosmological simulation with dark matter only ($l = 7
h^{1} Mpc$, $N = 1024^3$ particles, a mass resolution of $3.4\times
10^4 M_{\odot}$ and a force resolution of $0.16 h^{1}$ kpc), which we
use to construct halo catalogs and their merging histories. We then
follow the chemical enrichment of the gas within those halos, coupling
analytical recipes for cooling and star formation (plus subsequent
metal production) with a wind outflow model that assumes propagation
of metals out of Population II galaxies with a fiducial velocity of
60 km s$^{-1}$. Halos that are flagged as chemically pristine and meet the
requirement for cooling via either molecular hydrogen (in presence of
a self-consistent Lyman-Werner background) or Ly$\alpha$, are assumed
to be able to form metal-free stars. The minimum halo mass for cooling is shown in Fig.~1 of
\citet{Trenti et al. 09}, and is not repeated here. At $z<13.4$ it corresponds
to a dark matter halo with virial temperature $T_{vir}=10^4$ K.

As discussed in \citet{Stiavelli & Trenti 10}, the star forming metal-free
halos at $z\lesssim 15$ prefer low-bias regions, and form
preferentially away from larger Population II galaxies which would
otherwise pollute them via wind enrichment. The predictions, for the
formation rate of metal-free star-forming halos from our model are
shown in Fig.~4 of \citet{Trenti et al. 09}, and agree with similar
studies, for example those carried out by \citet{Tornatore et al.} if
one assumes in our model a relatively low  efficiency of formation of
Population III stars, for a total of a few $10^2\ M_{\odot}$ to a few
$10^3\ M_{\odot}$ per halo (containing $\sim 10^{7}\ M_{\sun}$ in
baryons). However, given the uncertainties in the processes that lead to the
formation of metal-free stars, we here explore a much wider range of star
formation efficiencies, as discussed in Sect.~\ref{SEM}.

\begin{figure}
\includegraphics[width=84mm]{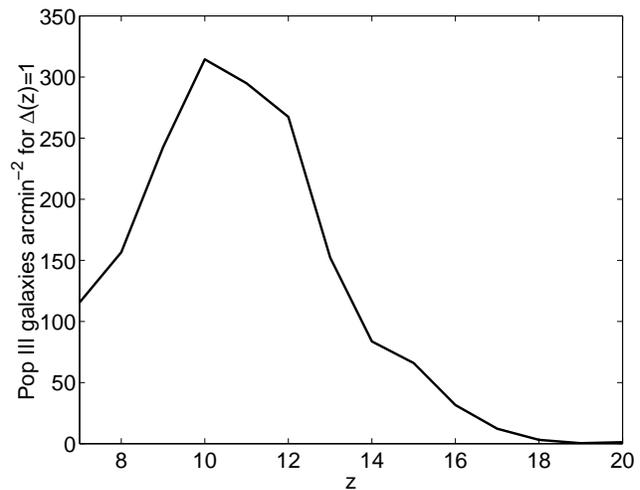}
\caption{The predicted number of pop III galaxies per arcmin$^2$ and unit redshift in an unlensed field, assuming that these galaxies lose their pop IIIs identities after $\tau_\mathrm{popIII}=10^7$ yr (see Sect.~\ref{SEM} for a more thorough description of this parameter).}
\label{popIIIs_per_area}
\end{figure}

In Fig.~\ref{popIIIs_per_area}, we plot the number of pop III galaxies per arcmin$^2$ and unit redshift throughout the range $z=7$--20, assuming that such galaxies lose their pop IIIs identities after $\tau_\mathrm{popIII}=10^7$ yr (see Sect.~\ref{SEM} for a more thorough description of this parameter). The surface number densities of pop III galaxies peak at $z\approx 10$, but remain high ($\gtrsim 100$ arcmin$^{-2}$) throughout the interval $z\approx 7$--14, which is sufficient to expect lensing surveys to be able to catch a few (as previously pointed out by \citealt{Stiavelli & Trenti 10}). To see why, consider the following example. The image-plane area for which $z\approx 10$ sources are expected to attain magnifications $\mu=10$--30 (with an average magnification $\overline{\mu}\approx 17$) is $\approx 2$ arcmin$^2$ in the J0717 field shown in Fig.~\ref{J0717}, whereas the corresponding area for $\mu=100$--300 ($\overline{\mu}\approx 160$) is $\approx 0.2$ arcmin$^2$. A rough estimate of the number of $z\approx 10$ pop III galaxies with $\mu=10$--30 in the J0717 field would then yield 300 objects arcmin$^{-2} \times 2$ arcmin$^2$ $\times (1/17) \approx 35$, whereas the corresponding number of galaxies with $\mu=100$--300 would give 300 arcmin$^{-2}\times 0.2$ arcmin$^2$ $\times (1/160) \approx 0.38$. Hence, the surface number densities of pop III galaxies should be sufficiently high to allow detection of objects with magnifications up to $\mu \lesssim 100$ in this field. 

The total masses of these pop III galaxy halos are in the range $\sim 10^7$--$10^8\ M_\odot$, with likely baryon masses an order of magnitude lower. Since {\it JWST} surveys of unlensed fields will be unable to detect pop III galaxies with stellar population masses below $\sim 10^5\ M_\odot$ \citep{Zackrisson et al. b}, a fraction $\gtrsim 0.01$--0.1 of these baryons must be converted into pop III stars to push them above the detection threshold. In a survey of a lensed field behind J0717, the required fraction can potentially be lower.

\subsection{Gravitational lensing}
\label{lensing}
In the process of producing mock samples of lensed pop III galaxies, simulated pop III halo catalogs (see Sect.~\ref{halos}) are projected through J0717 magnification maps. For this purpose, we use the mass (and magnification) model published in \citet{Zitrin et al. a}, which can be consulted for further details. The adopted image-plane resolution of these maps is chosen identical to the {\it HST/ACS} resolution, with 0.05 arcsec pixel$^{-1}$. Since this cluster shows uniquely high magnifications, and over a large area, we work with a 100 times better resolution (0.005 arcsec pixel$^{-1}$) in the source plane, so that magnifications of up to $\mu=100$ can be credibly tracked. While this paper was in preparation, \citet{Limousin et al.} presented new spectroscopic data on a number of lensed background objects in the J0717 field. Even though we have not revised our model or calculations to take advantage of these new data, their likely impact is deemed to be minor (see Sect.~\ref{uncertainty_section}).

To facilitate the lensing calculations, the pop III halo catalogs are projected onto source planes with thickness $\Delta z=1.0$, centered on $z=7$, 8, 9... 15. While the simulations predict a few pop III halos at even higher redshifts, such objects are out of bounds of the surveys considered, and are therefore not treated any further. The simulation boxes are comparable to the {\it HST} and {\it JWST} fields of view in the plane of the sky, but much too small in the line of sight direction to fill up the light cone volume between two adjacent source planes. Several boxes are therefore randomly rotated and stacked to produce each source plane catalog. This procedure admittedly scrambles some of large-scale clustering of pop III galaxies, but since these objects display very little clustering \citep{Stiavelli & Trenti 10}, any bias resulting from this is likely to be small. If anything, the final results should be conservative, since the probability of having rare, unusually massive pop III halos in a given lens plane becomes smaller when simulation cubes are recycled, compared to the case where the entire line of sight volume is self-consistently simulated.

For practical reasons, our lensing scheme necessarily involves a number of simplifications. Magnifications $\mu>100$ are not considered, since the resolution limits of our lens and source maps makes the exact magnification difficult to assess for such cases. While our lensed object maps typically include a few images of objects at $z\approx 7$--15 in this magnification range, all such images are excluded from further analysis. Moreover, all pop III galaxies are treated as point sources. A more realistic treatment could imply that different parts of the sources exhibit slightly different magnifications. Since most of the light from pop III galaxies may stem from nebular emission rather than direct star light \citep[e.g.][]{Schaerer a,Inoue,Zackrisson et al. b}, and since the nebulae are likely to extend over a larger area than the star-forming regions in these objects, differential magnification could alter the effective ratio of nebular to stellar emission in sources seen through J0717. The exact effect of this is difficult to assess at the current time -- the spatial extent of pop III galaxies likely depends on the ages, stellar initial mass function, halo masses, star formation efficiencies and Lyman continuum escape fractions of these objects, and simulations covering the full parameter space are not yet available. However, unless these pop III galaxies are larger than the lensed $z\approx 7$ galaxies detected so far \citep[e.g.][]{Zheng et al.,Bradley et al. a,Bradley et al. b}, differential magnification is unlikely to significantly distort the observables.

\subsection{Monte Carlo simulation of Pop III galaxy ages}
\label{MC}
Within each lens plane, ages are randomly assigned to every halo using a flat age distribution between $t=0$ and 21 Myr (the highest age considered in the \citealt{Trenti et al. 09} simulations). These ages represent the time since the onset of pop III star formation within each halo, and are used in conjunction with spectral synthesis models (see Sect.~\ref{SEM}) to derive the fluxes of the pop III galaxies. For every combination of model parameters (see below), this procedure is repeated over and over to generate a large number of mock pop III samples for statistical analysis.
 
\subsection{Spectral synthesis}
\label{SEM}
To generate the intrinsic spectral energy distributions (SEDs) of the pop III galaxies, we use the Yggdrasil spectral synthesis model \citep{Zackrisson et al. b} with \citet{Schaerer a} and \citet{Raiter et al.} stellar population spectra for pop III stars. The nebular contribution to the overall SED is computed using the photoionization code Cloudy \citep{Ferland et al.}, assuming a spherical geometry for the photoionized gas, a filling factor of 0.01 and a constant hydrogen density of $n(\mathrm{H})=100$ cm$^{-3}$. Various model grids (SEDs and fluxes in various filter systems) results are publicly available from the lead author's homepage\footnote{Yggdrasil model results available at: www.astro.su.se/$\sim$ez}.

When creating mock pop III samples, a number of different model parameters are considered. These are:
\begin{itemize}
\item {\bf The pop III stellar initial mass function (IMF)}. As in \citet{Zackrisson et al. b}, three different scenarios are considered: pop III.1, pop III.2 and a \citet{Kroupa} IMF. The pop III.1 option corresponds to a very top-heavy IMF with a power-law IMF ($\mathrm{d}N/d\mathrm{M}\propto M^{-\alpha}$) of slope $\alpha=2.35$ throughout the mass range 50--500$ M_\odot$. Pop III.2 corresponds to a slightly less extreme IMF, with log-normal shape, characteristic mass $M_\mathrm{c}=10\ M_\odot$, dispersion $\sigma=1.0$ and wings extending from 1--500 $M_\odot$. Finally, the \citet{Kroupa} option results in an IMF identical to that observed for more metal-rich stars in the local Universe.

\item {\bf Lyman continuum escape fraction, $f_\mathrm{esc}$.} A significant fraction of the ionizing radiation produced by the pop III stars may escape directly into the intergalactic medium \citep[e.g.][]{Johnson et al. 09}, with potentially strong effects on the resulting SED \citep{Zackrisson et al. b}. Here, we consider the two limiting cases  $f_\mathrm{esc}=0$ and $f_\mathrm{esc}=1$. The $f_\mathrm{esc}=0$ scenario results in a SED dominated by nebular emission  \citep[type A in][]{Zackrisson et al. b}, whereas 
$f_\mathrm{esc}=1$ (type C) results in a purely stellar SED. Because of their radically different SEDs, the colour signatures of these two types of pop III galaxies differ significantly, as discussed in Sect.~\ref{colours}. Since $f_\mathrm{esc}=0$ implies a significantly higher luminosity per stellar mass, such pop III galaxies are much easier to detect. 

\item {\bf Star formation history.} Two options are considered, either an instantaneous burst of star formation (giving a single-age population) or a constant star formation rate lasting $10^7$ yr. More prolonged star formation episodes appear unlikely, given the very strong feedback effects expected from pop III starbursts \citep[e.g.][]{Johnson et al. 09}.

\item {\bf Typical star formation efficiency (SFE), $\epsilon$.} Here, the SFE is defined as the fraction of the baryons within the halo that is converted into stars throughout the duration of the pop III starburst (see above). In our model, the luminosities of pop III galaxies simply scale with this parameter.  For simplicity, we assume all halos to have a baryon fraction equal to the cosmic average $\Omega_\mathrm{bar}/\Omega_\mathrm{M}$, which results into the following relation between SFE and mass in pop III stars:
\begin{equation}
M_\mathrm{popIII}=\epsilon M_\mathrm{halo}\frac{\Omega_\mathrm{bar}}{\Omega_\mathrm{M}}.
\end{equation}
It should be noted, however, that many simulations predict that halos with masses in the $\sim 10^7$--$10^8\ M_\odot$ range may have their baryon fractions reduced by factors of a few due to radiative feedback from surrounding structures \citep[e.g.][]{Mesinger & Dijkstra,Ricotti et al.}. In a recent paper, \citet{Safranek-Shrader et al.} present simulations which suggest that the SFE of a pop III galaxy in a $3\times10^7\ M_\odot$ halo at $z\approx 12$ may be $\log_{10}\epsilon\approx -3$. We therefore adopt this as our fiducial value, and investigate the circumstances under which pop III galaxies with this SFE would be detectable with the JWST. Given the substantial uncertainties still attached to the typical SFE, we will also consider a wider range of values (up to $\log_{10}\epsilon=0$) when discussing the limits and detection prospects relevant for the HST. This covers the SFE range predicted in the simulations by \citet{Wise & Cen} ($\log_{10} \epsilon \approx - 1$ to $-3$) for a $M\sim 10^8 \ M_\odot$ halo and the default value ($\log_{10} \epsilon \approx - 2.2$) assumed by \citet{Pawlik et al.} for their pop III galaxies.

\item {\bf Lifetime of distinct pop III spectral signatures, $\tau_\mathrm{popIII}$}. Even though the ejecta from the first pop III supernovae may take $\sim 10^7$--$10^8$ yr to cool sufficiently to be used in the subsequent formation of pop II and pop I stars \citep{Greif et al. 07,Ritter et al.}, the timescale during which pop III galaxies display unique spectral signatures may be much shorter. In principle, $\tau_\mathrm{popIII}$ depends on the spectral signature considered, the IMF, $f_\mathrm{esc}$ and the star formation history. For instantaneous-burst models, and the colour signatures considered by \citet{Zackrisson et al. b,Zackrisson et al. c}, $\tau_\mathrm{popIII}$ is $\lesssim 10^7$ yr, and may be as short as $\approx 3$ Myr if the IMF is extremely top-heavy (pop III.1) or even shorter if feedback quickly removes the gas from these systems \citep{Johnson et al. 09}. This parameter is used to reject pop III galaxies with ages in excess of $\tau_\mathrm{popIII}$ from our final mock samples. An effective age truncation may, however, set in even sooner if the actual lifetime of a given pop III galaxy turns out to be lower than $\tau_\mathrm{popIII}$. In the case of $\tau_\mathrm{popIII}=10^7$ yr, this happens for our instantaneous-burst pop III.1 galaxies, which remain luminous for just $\approx 3$ Myr before fading from sight. 

We adopt $\tau_\mathrm{popIII}=10^7$ yr as our default value, but also consider $\tau_\mathrm{popIII}$ as short as $10^6$ yr. A more extended star formation history could potentially allow for a higher $\tau_\mathrm{popIII}$, but given our maximum burst length of $10^7$ yr (see above), it is unlikely to be considerably longer than our default value. Moreover, the exact value of $\tau_\mathrm{popIII}$ turns out to be one of the least critical parameters in our simulations (see Sect.~\ref{secondary_parameters}). 
\end{itemize}

All pop III fluxes are computed under the assumption of zero dust extinction. The flux at wavelengths shortward of -- and including -- Ly$\alpha$ is moreover assumed to be zero to reflect the high level of absorption in the neutral intergalactic medium at the epochs considered ($z\geq7$). Intrinsically, the Ly$\alpha$ emission line is expected to be one of the strongest features in the spectra of pop III galaxies with nebular emission, and could give rise to distinct colour signatures \citep{Pello & Schaerer,Zackrisson et al. c}. While a certain fraction of the Ly$\alpha$ photons could potentially be transmitted through the intergalactic medium (IGM) at $z\approx 7$--8 due to outflows, source clustering and patchy reionization \citep{Dijkstra et al.,Dayal & Ferrara}, a high Ly$\alpha$ transmission fraction would require very fortunate sightline and can most likely apply only to a small fraction of pop III galaxies in our mock samples. For this reason, we conservatively assume the flux of the Ly$\alpha$ line to be zero, even though this may not always be the case. 

\subsection{{\it HST} and {\it JWST} surveys}
\label{survey_info}
The fluxes of pop III galaxies are predicted in two different filter systems: the {\it HST/WFC3} filters used by the Cluster Lensing And Supernova survey with Hubble \citep[{\it CLASH};][]{Postman et al.}, and the {\it JWST} filters relevant for a hypothetical {\it JWST} survey of the J0717 cluster.

{\it CLASH} is a 524-orbit multi-cycle treasury program targeting 25 galaxy clusters (including J0717) in a total of 16 filters with the {\it ACS} and {\it WFC3} instruments. The {\it CLASH} sample is larger and less biased than previous space-based cluster lensing surveys, with minimal lensing-based selection that favors systems with overly dense cores. Specifically, twenty {\it CLASH} clusters are solely X-ray selected to be massive ($kT > 5$ keV; $5-30\times 10^{14}\ M_\odot$) and, in most cases, dynamically relaxed. Five additional clusters were selected for their lensing strength (large Einstein radii or possibly high-magnification; to further quantify the lensing bias on concentration measurements, and yield some of the highest-resolution dark matter maps; see \citealt{Postman et al.} for further details). For objects at $z\geq 7$ (the range relevant for our simulations), detections are expected in at most 6 filters (F850LP, F105W, F110W, F125W, F140W and F160W), due to IGM absorption at wavelengths shortward of Ly$\alpha$ ($\lambda \geq 9700$ \AA{} at these redshifts). The 10$\sigma$ (5$\sigma$) detection limits in these filters are expected to be $m_{850}=26.0$ (26.7), $m_{105}=26.6$ (27.3), $m_{110}=27.0$ (27.8), $m_{125}=26.5$ (27.2), $m_{140}=26.7$ (27.4), $m_{160}=26.7$ (27.5) AB magnitudes \citep{Postman et al.}. The field of view of {\it HST/WFC3} (2.3\arcmin $\times$ 2.1\arcmin) is too small to cover the entire J0717 field (Fig.~\ref{J0717}; $4\times 4$ arcmin), and when simulating the {\it CLASH} source counts we therefore just analyze an area of 4.08 arcmin$^2$ towards the centre of the field, in rough agreement with how the {\it CLASH} observations of this cluster have been carried out. This region is $\approx 15\%$ smaller than the full {\it HST/WFC3} field of view, due to a rotation of the camera during the different exposures.
\begin{figure*}
\includegraphics[scale=0.46]{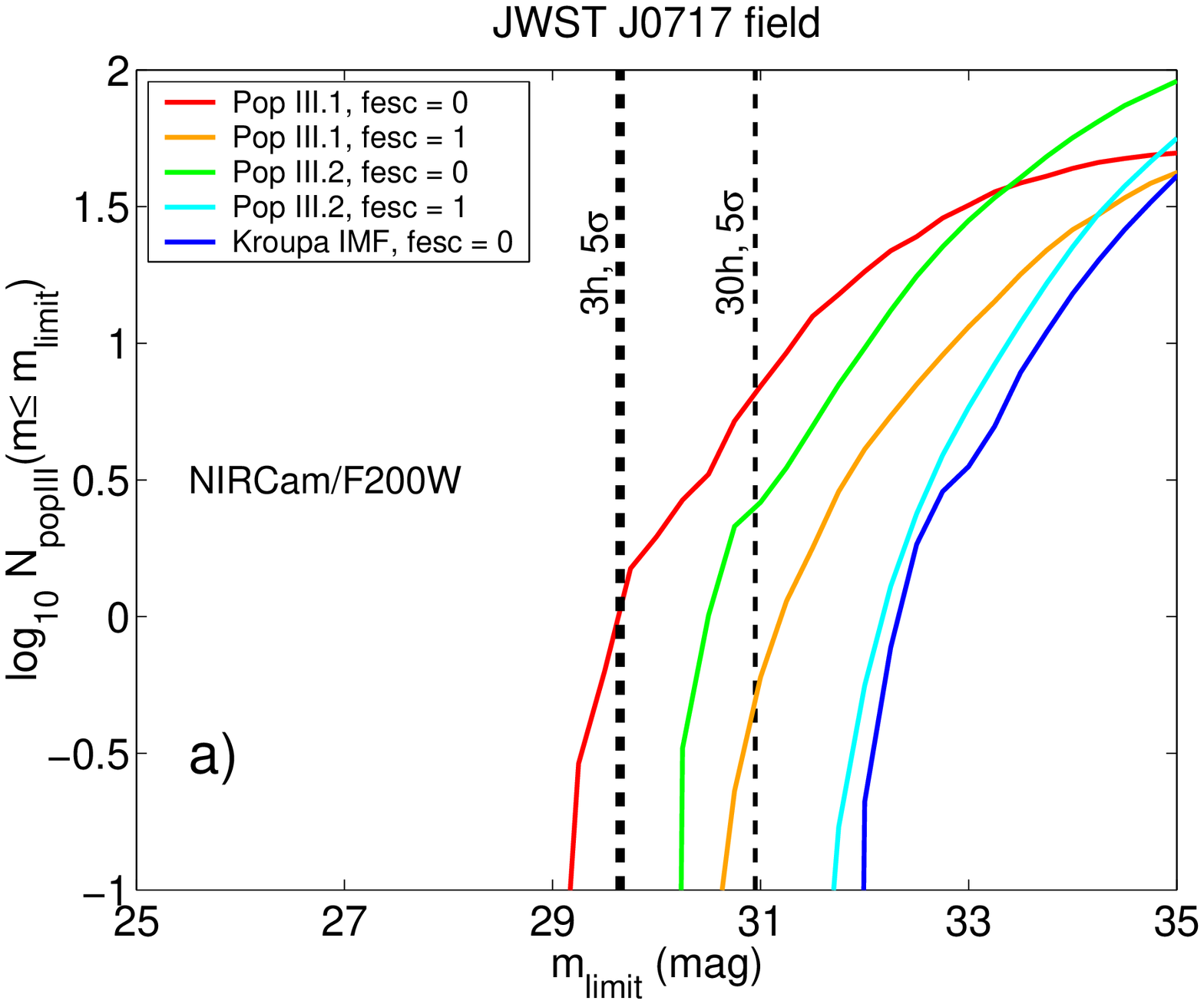}\includegraphics[width=84mm]{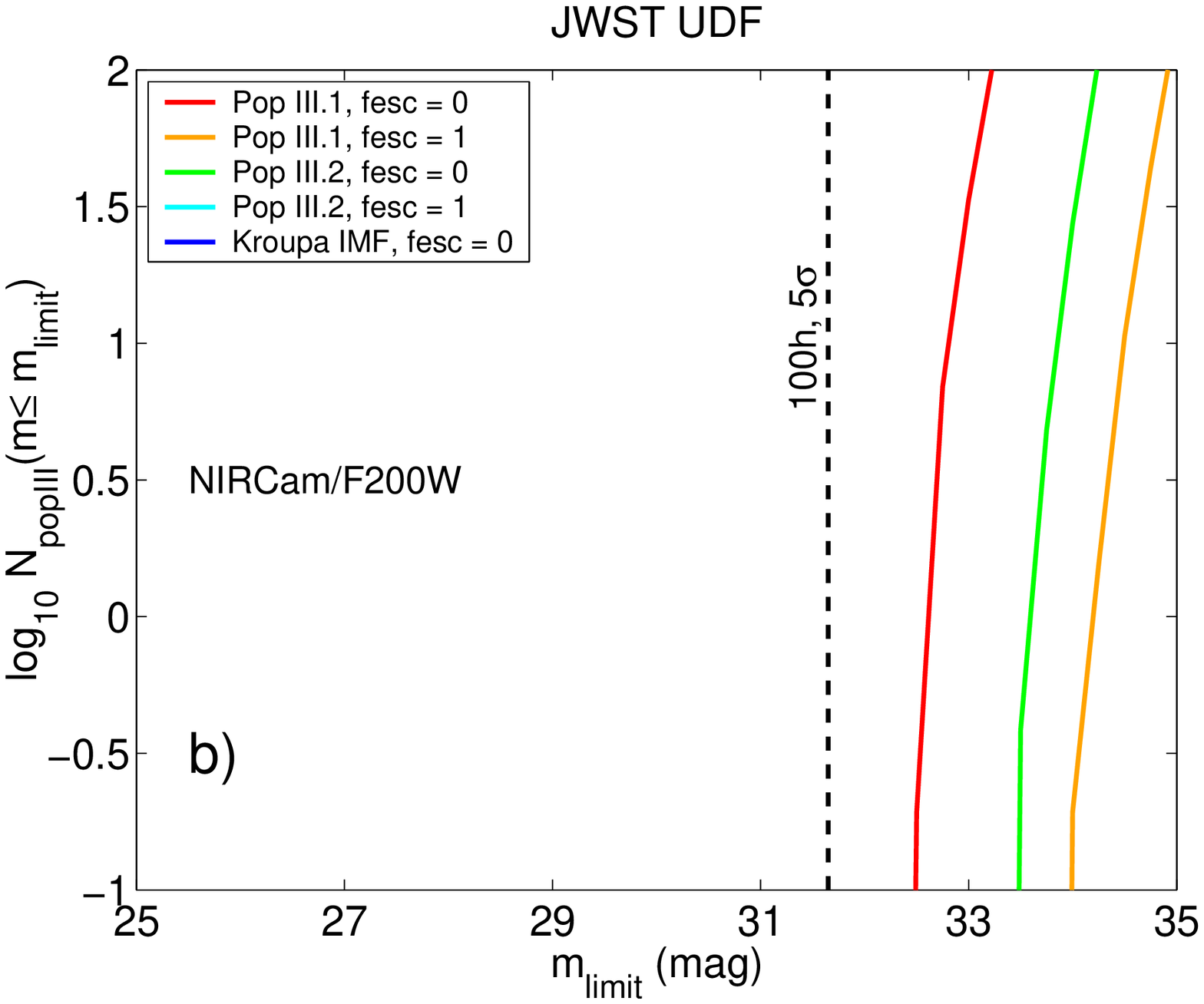}\\
\caption{Predicted number of lensed pop III galaxy images as a function of {\it JWST}/F200W magnitude limit in the
{\bf a)} J0717 field and {\bf b)} in an unlensed UDF. The coloured lines represent pop III galaxy models with the fiducial star formation efficiency ($\log_{10}\epsilon=-3$) and different combinations of IMF and Lyman continuum escape fraction: pop III.1, $f_\mathrm{esc}=0$ (red); pop III.1, $f_\mathrm{esc}=1$ (orange);  pop III.2, $f_\mathrm{esc}=0$ (green); pop III.2, $f_\mathrm{esc}=1$ (cyan); \citet{Kroupa}, $f_\mathrm{esc}=0$ (blue). All models assume an instantaneous burst of star formation and $\tau_\mathrm{popIII}=10^7$ yr. The dashed vertical lines indicate the $5\sigma$ detection thresholds expected for different exposure times. Certain varieties of $\log_{10}\epsilon=-3$ pop III galaxies are in principle detectable in the J0717 field (red, orange and green lines), whereas all such models lie well below the detection threshold of the UDF.} 
\label{Palantirfig}
\end{figure*}

Our hypothetical {\it JWST} multiband imaging survey is assumed to target J0717 with either $t_\mathrm{exp}=3$ h or $t_\mathrm{exp}=30$ h exposures per filter. The resulting 10$\sigma$ detection limits in all of the {\it JWST/NIRCam} and in the two bluest {\it JWST/MIRI} broadband filters are listed in Table \ref{detection_limits}. At similar wavelengths ($\approx 1.50\ \mu$m), this survey reaches $\approx 2.0$ and 3.3 mag (for $t_\mathrm{exp}=3$ h and $t_\mathrm{exp}=30$ h exposures respectively) fainter than the 10$\sigma$ limits of {\it CLASH}. Of course, since {\it JWST} allows for observations at longer wavelengths than is possible with {\it HST}, the range of detectable object redshifts is also much larger. The field of view of {\it NIRCam} is $\approx$ 9.3 arcmin$^2$, divided into two $2.16 \times 2.16 $ arcmin$^2$ modules with a $\approx 50$ arcsec gap in between. When simulating the {\it NIRCam} source counts in the J0717 field, we have therefore assumed a camera rotation that places the field of view diagonally across the cluster (covering the lower left to upper right regions in Fig.~\ref{J0717}), thereby capturing most of the lensing caustics where the highest magnifications are expected. 

The pop III galaxy number counts predicted in the J0717 field for the $t_\mathrm{exp}=3$ h and 30 h detection thresholds are compared to what one would expect from ultra-deep field (UDF) style {\it JWST} observations of an unlensed field. In the latter case, the exposure times are assumed to be $t_\mathrm{exp}=100$ h per filter, in rough agreement with the deepest {\it HST} observations carried out so far \citep{Beckwith et al.}. The computational scheme used to simulate the {\it JWST} UDF is identical to the one outlined above, except that the projection through J0717 magnification maps is omitted. This means that we implicitly assume the magnification across the UDF to be unity. One may argue that there is no such thing as a completely unlensed field at high redshift, since foreground galaxies at $z\approx 1$--2 may magnify objects at $z>6$ \citep{Wyithe et al.}. For this reason, our results for the UDF should be considered somewhat conservative. Since we do not consider additional lensing boosts from intermediate-redshift galaxies located behind the J0717 cluster, a similar effect may be present in the J0717 field as well. 

In general, surveys of lensed fields can always reach deeper than blank-field surveys -- and in just a fraction of the exposure time. However, blank-field surveys may still be more successful, depending on the luminosity function and the clustering properties of the high-redshift objects targeted \citep[e.g.][]{Maizy et al.,Bouwens et al. 09}. High-redshift objects that are highly clustered or exhibit very low surface number densities may be impossible to detect behind a single lensing cluster, even if the magnification is sufficient to lift the intrinsic fluxes of this class of objects above the survey detection threshold. Luckily, the surface number densities of pop III galaxies is high (Fig.~\ref{popIIIs_per_area}) and their clustering weak \citep{Stiavelli & Trenti 10}. 

\section{Detecting pop III galaxies with {\it JWST}}
\label{Palantir}
In Fig.~\ref{Palantirfig}a, we plot the expected number of lensed pop III galaxy images $N_\mathrm{popIII}$ in a single $\it JWST/NIRcam$ J0717 field, as a function of detection threshold in the F200W filter (the most sensitive NIRCam passband, allowing detections of objects up to $z\approx 15$). The coloured lines represent a set of different pop III galaxy models, all with SFE $\log_{10} \epsilon =-3$; (see Sect.\ref{SEM} for a discussion). The dashed vertical lines indicate the $5\sigma$ detection thresholds corresponding to exposure times of $t_\mathrm{exp}=3$ and 30 h per pointing. Detections of pop III galaxies are deemed possible whenever a model track ventures brightward of one of these limits (to the left in Fig.~\ref{Palantirfig}) and simultaneously predicts $\gtrsim 1$ image ($\log_{10} N_\mathrm{popIII}(m\leq m_\mathrm{limit})\gtrsim 0$). The corresponding results for the {\it JWST} UDF are presented in Fig.~\ref{Palantirfig}b. Since an unlensed field probes a much greater background volume than a high-magnification field like J0717, the image counts are steeper in the UDF and reach higher maximum values, albeit at much fainter flux levels.  
As seen in Fig.~\ref{Palantirfig}a, pop III galaxies with $\log_{10} \epsilon =-3$ are detectable through {\it JWST} observations of a strong-lensing cluster like J0717, as long as the IMF is top-heavy (pop III.1 or pop III.2) and the escape of Lyman continuum photons into the IGM is minor (red and green lines in Fig.~\ref{Palantirfig}). Galaxies with a pop III.1 IMF (characteristic mass $\sim 100\ M_\odot$) are the easiest to detect, despite having the shortest lifetimes ($\approx 3$ Myr for an instantaneous burst). With this IMF, even pop III galaxies with $\log_{10} \epsilon =-3.5$ are potentially detectable, provided that $f_{esc}$ is low. This represents a hard limit -- if pop III galaxies typically attain SFEs lower than this, they will inevitably be beyond the reach of {\it JWST} at $z\gtrsim 7$, since ultra-deep observations of unlensed fields are predicted to fare even worse (Fig.~\ref{Palantirfig}b). In fact, none of our $\log_{10} \epsilon =-3$ pop III galaxy models are able to make it above the $t_\mathrm{exp}=100$ h detection threshold of the {\it JWST} UDF. When a similarly top-heavy IMF is adopted in realistic simulations of $\sim 10^8\ M_\odot$ halos, strong feedback causes substantial Lyman continuum leakage into the IGM, thereby violating the  $f_\mathrm{esc}=0$ assumption, even for SFEs as low as $\log_{10} \epsilon \approx -3$ to $-4$ \citep{Johnson et al. 09}. Even if $f_\mathrm{esc}=0$, one of our pop III.1, $\log_{10} \epsilon =-3$ scenarios does, however, remain marginally detectable in the J0717 field (orange line in Fig.~\ref{Palantirfig}a). 
\begin{table}
\caption{{\it JWST} 10$\sigma$ detection limits (AB magnitudes) for 3h, 30h and 100h exposures per filter based on the JWST exposure time calculator, version 1.4. The corresponding 5$\sigma$ limits lie $\approx 0.75$ mag faintward of these thresholds.}
\begin{tabular}{@{}llllll@{}}
\hline
Instrument & Filter & $\lambda$ ($\mu$m) & $m_{3\mathrm{h}}$ & $m_{30\mathrm{h}}$ & $m_{100\mathrm{h}}$\\
\hline
NIRCam & F070W  & 0.70 & 28.0 & 29.4  & 30.1\\
       & F090W  & 0.90 & 28.5 & 29.8  & 30.4\\
       & F115W  & 1.15 & 28.6 & 29.9  & 30.5 \\
       & F150W  & 1.50 & 28.7 & 30.0  & 30.6 \\
       & F200W  & 2.00 & 28.9 & 30.2  & 30.9 \\
       & F277W  & 2.77 & 28.9 & 30.1  & 30.8 \\       
       & F356W  & 3.56 & 28.7 & 30.0  & 30.6 \\
			 & F444W  & 4.44 & 28.2 & 29.4  & 30.0 \\
MIRI   & F560W  & 5.60 & 25.7 & 27.0  & 27.6 \\
       & F770W  & 7.70 & 25.1 & 26.4  & 27.0 \\
\hline
\label{detection_limits}
\end{tabular}
\end{table}

The latest simulations do admittedly not give much support for an IMF as extreme as our pop III.1 scenario \citep[e.g.][]{Stacy et al.,Clark et al.,Greif et al. 11a,Greif et al. 11b,Hosokawa et al.}. The pop III.2 case (characteristic mass $\sim 10\ M_\odot$) is in better agreement with current predictions, and $\log_{10} \epsilon =-3$ galaxies of this type should be detectable in the J0717 field provided that $f_\mathrm{esc}$ is low (green line in Fig.~\ref{Palantirfig}a). Pop III galaxies with an IMF resembling that of galaxies in the local Universe \citep{Kroupa} remain undetectable at this SFE, regardless of the typical $f_\mathrm{esc}$ (blue line in Fig.~\ref{Palantirfig}a).

The mean redshift of pop III galaxies with non-zero {\it JWST}/F200W fluxes in our mock samples is $\bar{z}\approx 11$, but the mean redshifts of objects sufficiently bright to be detected above the {\it JWST} detection limits are always lower than this. In general, mock samples based on parameters that generate intrinsically faint pop III galaxies (e.g. a Kroupa IMF, high $f_\mathrm{esc}$ and low SFE) have lower mean redshifts than mock samples with intrinsically bright objects (e.g. top-heavy IMF, low $f_\mathrm{esc}$ and high SFE). The mean magnifications $\bar{\mu}$ of all images with non-zero F200W fluxes in our {\it NIRCam} J0717 field are $\bar{\mu}\approx 10$, but since high magnifications are required to make intrinsically faint objects detectable, the mean magnifications of images above the thresholds are also higher -- especially so for samples with intrinsically faint pop III galaxies. For these reasons, the mean redshifts and magnifications of $\log_{10} \epsilon =-3$ pop III galaxies that make it above the {\it JWST} detection thresholds in the J0717 field have $\bar{z}\approx 9$ and $\bar{\mu}\geq 40$.

\section{Detecting pop III galaxies with {\it HST}}
\label{HST_section}
\begin{figure*}
\includegraphics[width=84mm]{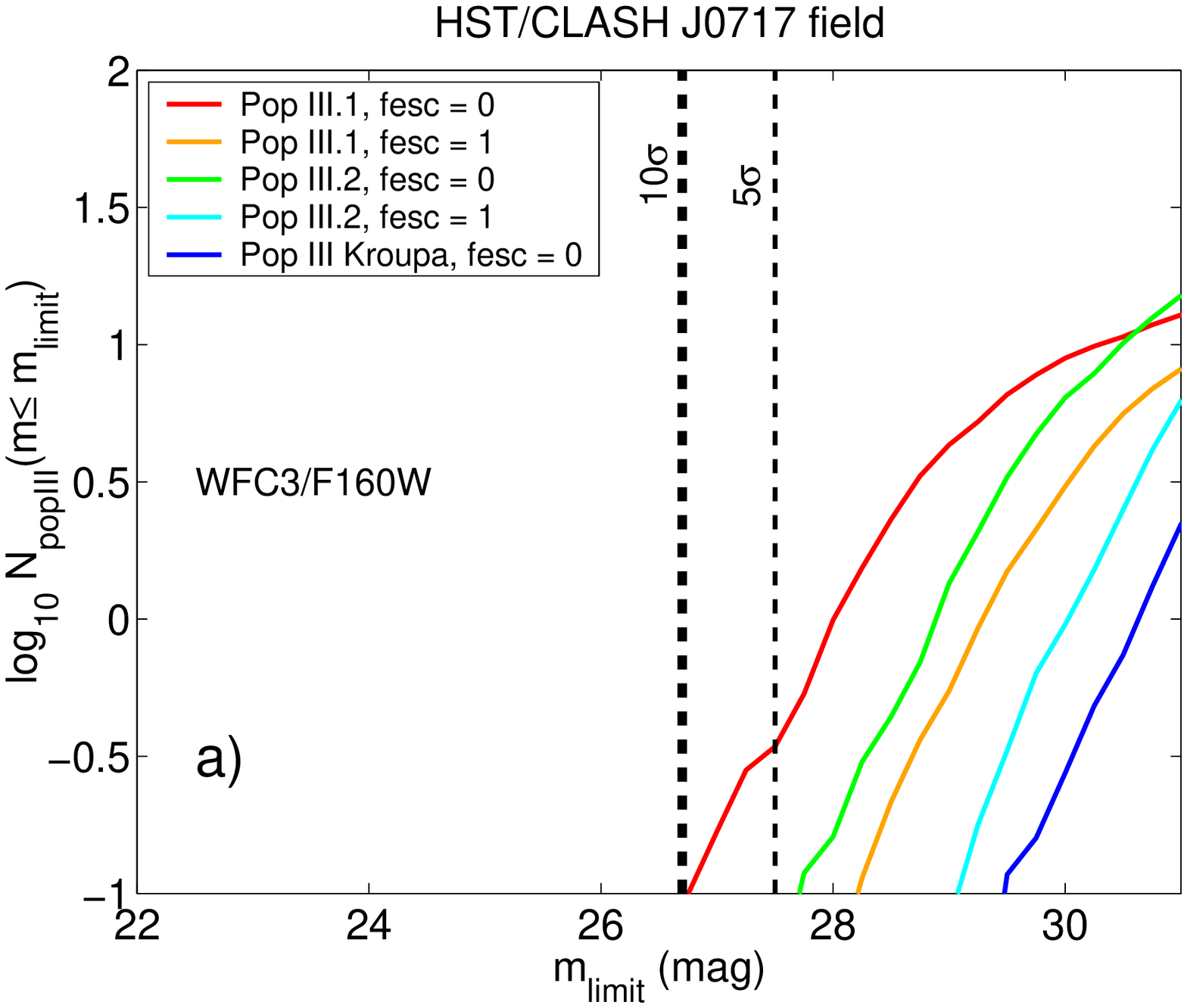}\includegraphics[width=84mm]{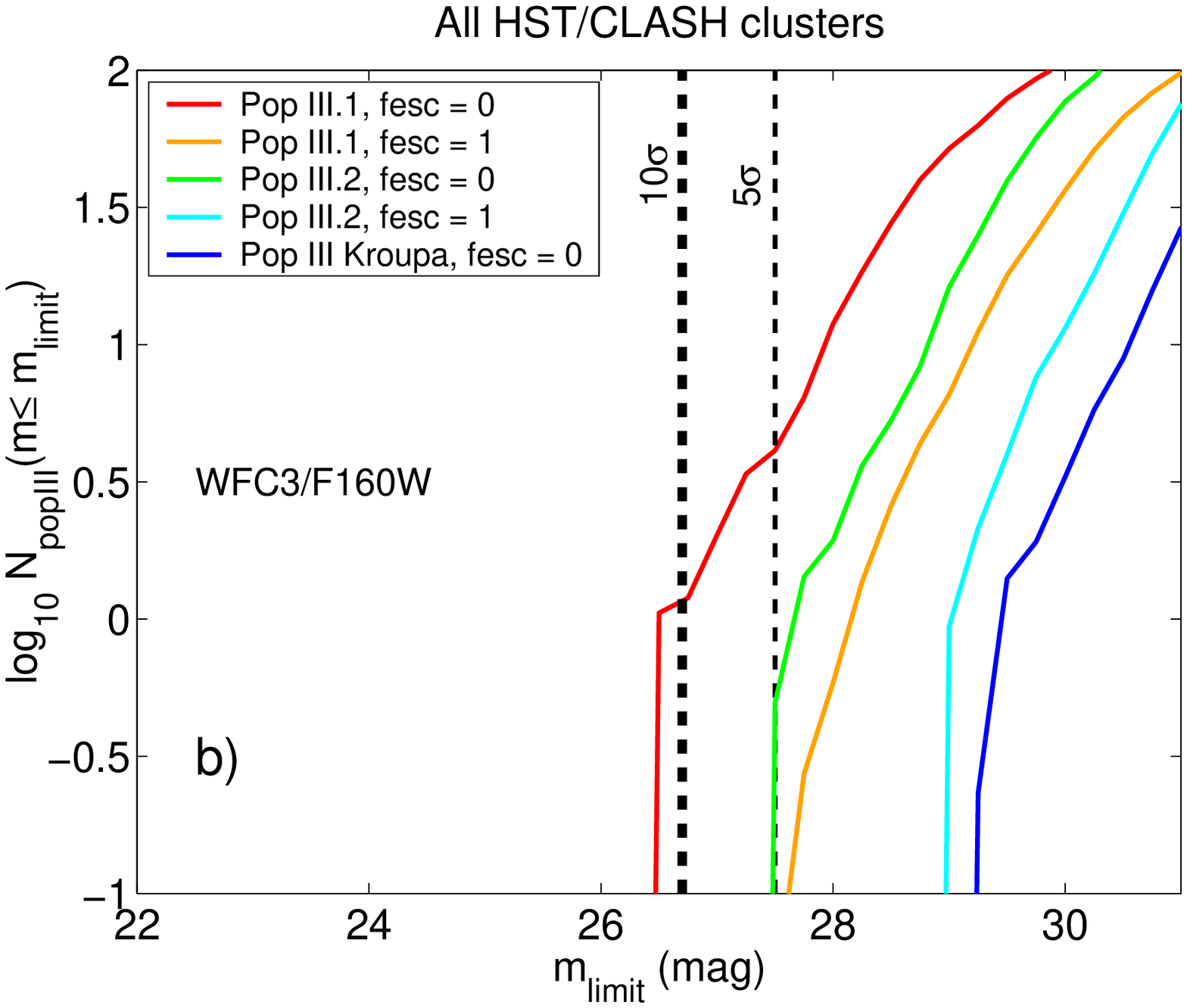}
\caption{Predicted number of lensed pop III galaxy images as a function of {\it HST}/F160W magnitude limit in the 
{\bf a)} CLASH J0717 field and the {\bf b)} full CLASH sample. The coloured lines represent pop III galaxy models with star formation efficiency $\log_{10}\epsilon=-2$ and different combinations of IMF and Lyman continuum escape fraction: pop III.1, $f_\mathrm{esc}=0$ (red); pop III.1, $f_\mathrm{esc}=1$ (orange);  pop III.2, $f_\mathrm{esc}=0$ (green); pop III.2, $f_\mathrm{esc}=1$ (cyan); \citet{Kroupa}, $f_\mathrm{esc}=0$ (blue). All models assume an instantaneous burst of star formation and $\tau_\mathrm{popIII}=10^7$ yr. The dashed vertical lines indicate the $10\sigma$ (thick dashed) and $5\sigma$ (thin dashed) detection thresholds of the CLASH survey. While no $\log_{10}\epsilon=-2$ pop III galaxies are expected in the J0717 data alone ({\bf a}), a small number of $\log_{10}\epsilon=-2$ pop III.1 or pop III.2 galaxies with $f_\mathrm{esc}=0$ (red and green lines) may in principle turn up once the full CLASH data set is considered ({\bf b}).}
\label{CLASHfig}
\end{figure*}

Pop III galaxies at $z\geq 7$ with star formation efficiency $\log_{10}\approx -3$ \citep{Safranek-Shrader et al.} are right at the expected detection threshold of {\it JWST} (Sect.~\ref{Palantir}), and well beyond the reach of current telescopes. However, if the SFE were significantly higher ($\log_{10}\gtrsim -2$), objects of this type would in principle be detectable with the {\it HST} at $z\approx 7$--10. In the following, we explore the prospects of detecting pop III galaxies in the ongoing {\it HST/CLASH} survey of lensing clusters, and derive the SFE limits imposed by existing {\it HST} data on unlensed fields. 

\subsection{Detecting pop III galaxies in the {\it HST/CLASH} survey}
\label{CLASH}
In Fig.~\ref{CLASHfig}a, we present the average number of pop III galaxy images $N_\mathrm{popIII}$ predicted within the CLASH J0717 field as a function of detection threshold in the {\it HST/WFC3} F160W filter. Formally, the CLASH survey reaches somewhat deeper in the F110W filter (detection limit $m_{110}=27.0$ at $10\sigma$) than in F160W ($m_{160}=26.7$ at $10\sigma$), but since objects start to appear as dropouts in the F110W filter already at $z\gtrsim 8$, our simulations still predict more pop III galaxies in F160W. The pop III galaxy models plotted (coloured lines) are the same as those discussed in the context of {\it JWST} detection (Fig.~\ref{Palantirfig}), only shifted to a star formation efficiency an order of magnitude higher (to $\log_{10}\epsilon=-2$). Most of these models lie just below the detection thresholds in the CLASH J0717 field, and only pop III galaxies with a pop III.1 IMF and with negligible Lyman continuum leakage (red line) would be marginally detectable in these data. 

Since the J0717 cluster most likely exhibits the largest high-magnification area ($\approx 3.5$ arcmin$^2$ for $\mu>10$ at $z>6$) among known cluster lenses, no other single cluster in the CLASH survey is likely to do any better. However, given the size of the CLASH survey, the prospects of detecting a pop III galaxy in the full CLASH data set (25 clusters) are of course considerably better than when just the J0717 images are used. To assess the maximum number of pop III galaxies that may be detected in the full survey, we estimate the expected total area corresponding to high magnification ($\mu>10$) covered by the CLASH cluster sample. In doing so, we simply assume that the mass slope of the 20 relaxed, X-ray selected CLASH clusters will distribute around the mean value of $d\log \Sigma/d\log \theta\simeq -0.55$, where $\Sigma$ represents the surface mass density and $\theta$ an angle in the sky. This is the typical $d\log \Sigma/d\log \theta$ value seen for well-known relaxed (lensing) clusters such as Abell 1703, Abell 383, Abell 1689, Cl0024, among other examples. Since the majority of CLASH clusters is X-ray selected, so that their lensing power is not biased in that sense, we can assume that they would follow this (steeper) profile typical for relaxed clusters, which results in a more moderate magnification. These slope values, according to models of several similar relaxed lensing clusters as mentioned above, yield a high-magnification area of $\approx 1.55\pm0.4$ arcmin$^2$ per cluster on average, so that 20 such clusters should supply $\approx 31$ arcmin$^2$ of area with $\mu>10$. For the five (expected)  higher-magnification clusters we assume a shallower profile, as seen in our previous analysis of MACS0717 and MACS1149, or the full sample of 12 higher-$z$ MACS clusters \citep{Zitrin et al. b}. Average values therein are taken as representative for this part of the sample, with $\approx2$--2.5 arcmin$^2$ per cluster. In total, this yields a total area of $\approx 42$ arcmin$^2$ for the CLASH sample. This is $\approx 12$ times greater than the corresponding area in J0717 alone, which means that the predicted pop III galaxy number counts may be shifted upwards by approximately this factor once the full CLASH sample is considered. To illustrate the effects of this, Fig.~\ref{CLASHfig}b presents the image counts expected from the complete CLASH survey for the same $\log_{10}\epsilon=-2$ pop III galaxy models as presented for just the J0717 field in Fig.~\ref{CLASHfig}a. Two of the pop III galaxy models (pop III.1 and pop III.2 with $f_\mathrm{esc}=0$; red and green lines) are now shifted into the detectable range at $\log_{10} N_\mathrm{popIII}(m\leq m_\mathrm{limit})\gtrsim 0$. Of course, the caveats concerning these parameter combinations discussed in Sect.~\ref{Palantir} apply here as well.

The mean redshift of all pop III galaxies with non-zero fluxes in the {\it HST}/F160W filter is $\bar{z}\approx 10$, but in analogy with the case for {\it JWST}, the objects that actually appear above the detection thresholds tend do be at somewhat lower redshifts. For the models depicted in Fig.~\ref{CLASHfig}, the objects appearing on the brightward side of the detection thresholds have $\bar{z}\approx 8$--8.5. The mean magnifications $\bar{\mu}$ of these images is $\bar{\mu}\approx40$--60.

\subsection{Constraints set by the UV luminosity function}
\label{LF}
While many combinations of model parameter values (see Sect.~\ref{SEM}) produce pop III galaxy models with fluxes far below the detection thresholds of both the {\it JWST} (Fig.~\ref{Palantirfig}) and {\it HST} (Fig.~\ref{CLASHfig}), some mock samples with very high SFEs result in more detectable sources per surface area than observed in the {\it HST/WFC3} data on the Hubble Ultra Deep Field (HUDF) and the adjacent Great Observatories Origins Deep Survey (GOODS) fields. The parameter combinations producing such violations can therefore be ruled out. The upper limits on the typical SFE resulting from a comparison of the pop III luminosity functions (LFs) to the observational constraints on the LF at $z\approx 7$, 8 and 10 \citep{Bouwens et al. 11,Oesch et al.} are listed in Table.~\ref{limits}. For the very brightest pop III galaxy models, the resulting upper limit is $\log_{10}\epsilon\lesssim -1.5$, whereas for the very faintest models, the LF is unable to set any meaningful upper limits on the SFE.

\begin{figure*}
\includegraphics[width=84mm]{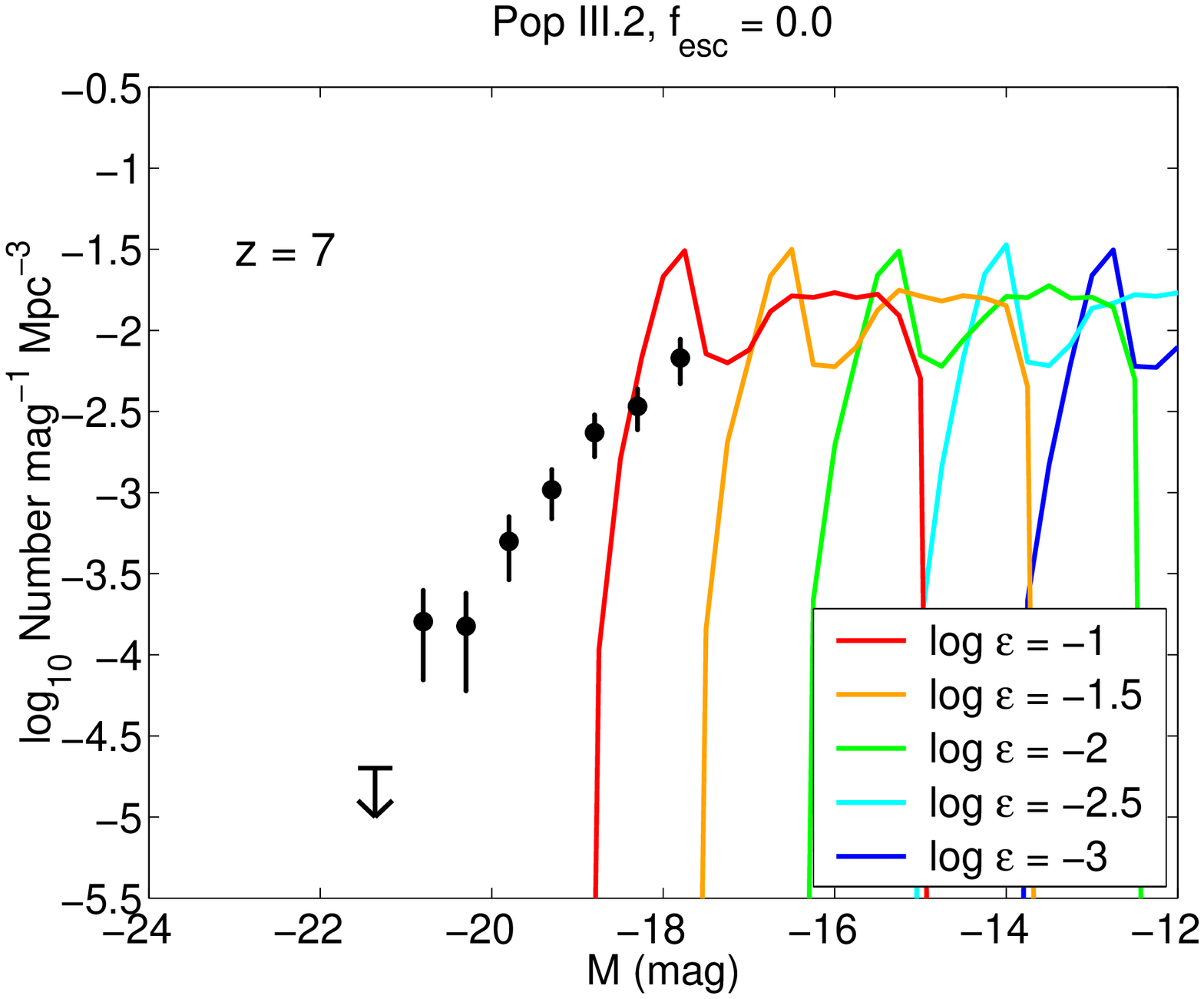}\includegraphics[width=84mm]{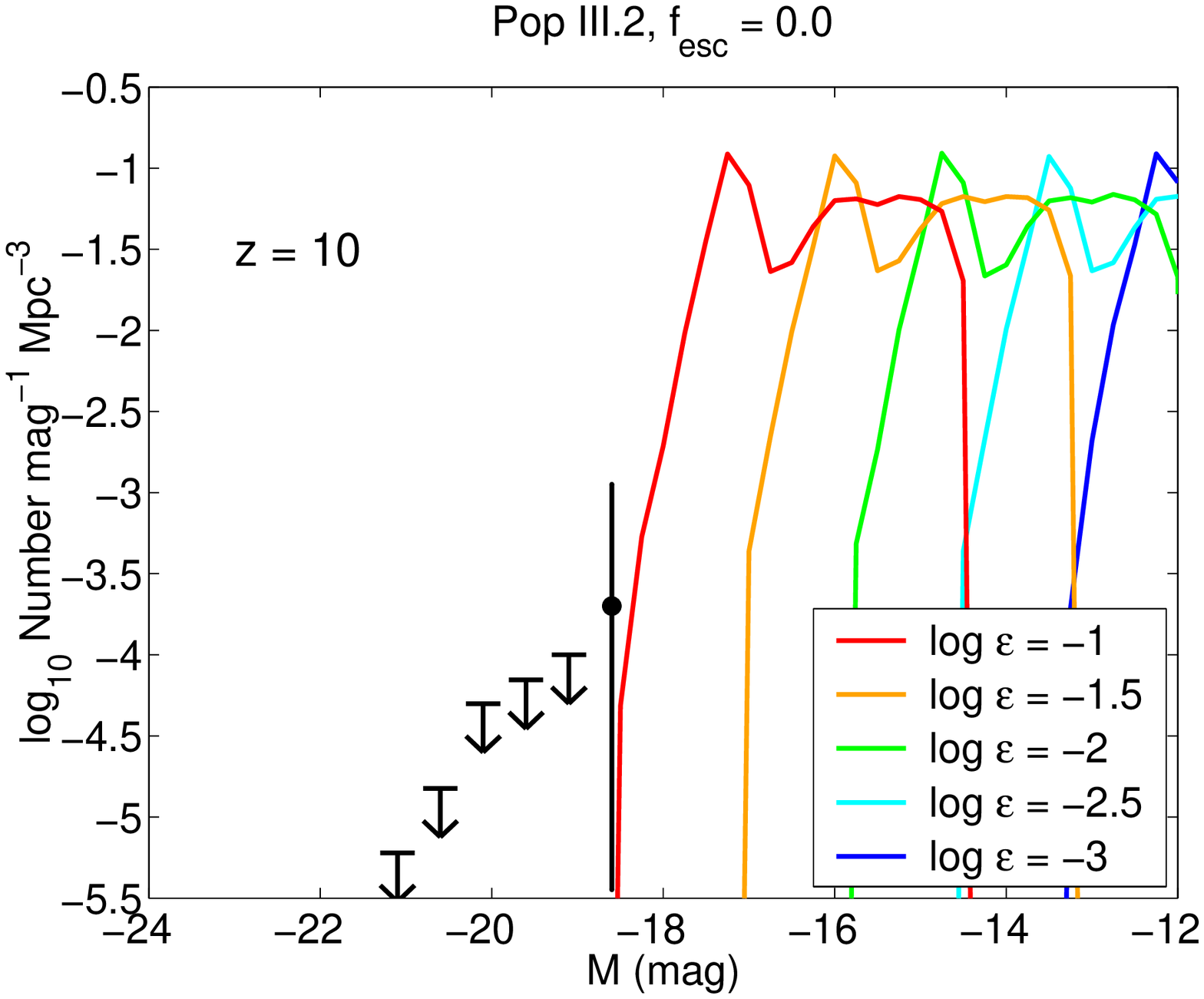}\\
\caption{Example of how the observed rest-frame ultraviolet luminosity function (LF) at high redshifts constrains the typical star formation efficiencies of pop III galaxies. The panels illustrate how the observed LFs \citep{Bouwens et al. 11,Oesch et al.} at $z\approx 7$ (left panel) and $z\approx 10$ (right panel) set upper limit to the star formation efficiency $\epsilon$ of pop III galaxies with a pop III.2 IMF, $f_\mathrm{esc}=0$,  $\tau_\mathrm{popIII}=10^7$ yr and an instantaneous burst of star formation. The different line colours represent different $\epsilon$ values. In this case, the $\log_{10} \epsilon = -1$ model (red line) overshoots the faintest bin of the observed LF at $z\approx 7$ (left panel) and is therefore ruled out, whereas models with $\log_{10} \epsilon = -1.5$, $-2.0$, $-2.5$ and $-3.0$ (orange, green, cyan and blue lines) remain viable. A more detailed analysis places the upper limit at $\log_{10} \epsilon\lesssim -1.2$ (in between the models represented by the red and orange lines). Adopting a model with $\epsilon$ close to this limit would, however, imply that a significant fraction of the faintest objects detected at $z=7$--10 in fact are pop III galaxies.} 
\label{LFfig_1}
\end{figure*}

The basic principle is illustrated in Fig.~\ref{LFfig_1} for an instantaneous-burst model with a pop III.2 IMF, $f_\mathrm{esc}=0$ and $\tau_\mathrm{popIII}=10^7$ yr. A star formation efficiency of $\log_{10}\epsilon=-1$ (red line) would overshoot the observed rest-frame ultraviolet LF at $z=7$ (left panel), and can therefore be ruled out, even though it is marginally consistent with the LF at $z=10$ (right panel). An SFE of $\log_{10}\epsilon=-1.5$ (orange lines) can on the other hand not be ruled out at any of these redshifts. In this case, a detailed calculation places the upper limit on the SFE at $\log_{10}\epsilon\leq-1.2$ (see Table ~\ref{limits}). This procedure does not in any way address the issue of whether pop III galaxies may already exist in current data \citep{Zackrisson et al. c}, but simply ensures that clearly unrealistic models are not considered any further. In fact, models located right at the SFE limits presented in Table.~\ref{limits} would be consistent with having a few pop III galaxies in current HUDF and GOODS samples at $z\approx 7$--10. In most cases, the strongest constraints are set by the LF at $z\approx 7$, since this is where the observed LF reaches the deepest in terms of intrinsic luminosity. This technique of using observed source counts to constrain the properties of pop III galaxies is similar in spirit to that used by \citet{Zackrisson et al. d} to impose limits on high-redshift supermassive dark stars. Such objects could potentially attain very high surface number densities, but in practice violate the observed LF unless the model parameters are tuned to make these objects exceedingly rare.

\begin{figure}
\includegraphics[width=84mm]{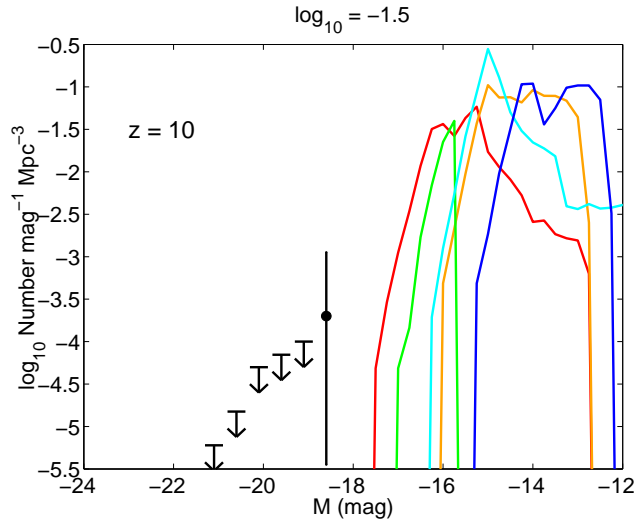}
\caption{Examples of how the predicted pop III galaxy luminosity function (LF) at $z \approx 10$ is shaped by the pop III galaxy model parameters. The differently coloured lines represent the LFs predicted in the case of a pop III.1 IMF with $f_\mathrm{esc}=1$ (red line); a pop III.2 IMF with $f_\mathrm{esc}=1$ (orange); a pop III.2 IMF with $f_\mathrm{esc}=0$ and $\tau_\mathrm{popIII}=10^6$ yr  (green); a pop III.2 IMF with $f_\mathrm{esc}=0$ and a $10^7$ yr burst (cyan); a pop III, \citet{Kroupa} IMF with $f_\mathrm{esc}=0$ (blue). The black symbols represent the observed rest-frame ultraviolet LF constraints at this redshift \citep{Oesch et al.}. All models assume $\log_{10} \epsilon = -1.5$ and remain viable once the LF constraints (Table ~\ref{limits}) have been applied. For this star formation efficiency, the overall LF is expected to undergo dramatic changes a few magnitudes below the current detection threshold due to the large volume densities of pop III galaxies at this redshift.} 
\label{LFfig_2}
\end{figure}

It is worth noting that many of the models that remain viable once the LF limits have been applied still predict sufficiently many pop III galaxies to dominate and alter the slope of the overall LF a few magnitudes below the current detection threshold. This is especially the case at $z\approx 10$, due to the large volume densities of pop III galaxies at this redshift. In Fig.~\ref{LFfig_2}, we demonstrate this by plotting the $z\approx 10$ LFs resulting from a number of different pop III galaxy models with fixed $\log_{10}\epsilon = -1.5$. Hence, if the typical pop III SFE were this high, future extensions of the LF to fainter magnitudes would be expected to detect an abrupt upturn.

Changing the IMF from a very top-heavy form (pop III.1) to a more moderately top-heavy version (pop III.2) or a perfectly normal IMF \citep{Kroupa} has the effect of increasing the lifetimes of the model galaxies, at the expense of the lowering their peak luminosities, thereby shifting the predicted pop III galaxy luminosity distribution upwards and to the right in these diagrams (compare red and orange lines in Fig.~\ref{LFfig_2}, or the orange line in Fig.~\ref{LFfig_1} to the blue line in Fig.~\ref{LFfig_2}). Adopting a more extended star formation history than an instantaneous burst has a very similar effect, since this also lowers the peak luminosity but increases the lifetime (compare the orange line in Fig.~\ref{LFfig_1} to the cyan line in Fig.~\ref{LFfig_2}). Since pop III galaxies typically attain their greatest luminosities at very low ages, adopting a $\tau_\mathrm{popIII}$ lower than $10^7$ yr primarily results in a truncation of the low-luminosity tail of LF, with a pronounced effect on the overall shape of the pop III LF but with little relevance for current LF measurements. 

For a number of reasons, the star formation efficiency limits presented in Table ~\ref{limits} should be considered approximate only. Whereas the absolute magnitudes plotted in Fig.~\ref{LFfig_1} and \ref{LFfig_2} are based on the same {\it HST/WFC3} filters as the observations (F125W for $z\approx 7$; F160W for $z\approx 8$ and $z\approx 10$), the mock sample data are based on simulations statistics within redshift bins of depth $\Delta z=1.0$ centered at $z=7$, 8 and 10, with all pop III galaxy magnitudes computed using the central redshift. This is admittedly a crude approximation, since the unusual SEDs of pop III galaxies may result in a redshift selection function that is very different from that of more mundane galaxies. This is particularly the case if $f_\mathrm{Ly\alpha}$ is non-zero for some of these galaxies \citep{Zackrisson et al. c}. Also, the results presented in Table ~\ref{limits} do not take the LF contribution from more mundane galaxies (which are likely to be far more abundant than pop III galaxies) into account, which makes the upper limits on the star formation efficiencies conservative. 

\begin{table}
\caption{Constraints on the star formation efficiency $\epsilon$ of pop III galaxies derived from the observed rest-frame UV luminosity functions at $z=7$--10}
  \begin{tabular}{@{}lcccc@{}}
  \hline
IMF  & $f_{esc}$ & Burst (yr) & $\tau_\mathrm{popIII}$ (yr) & $\log_{10}\max(\epsilon)$\\
\hline
Pop III.1	      & 0.0 & Inst & $10^7$ & $-1.6$ \\
Pop III.1	      & 0.0 & Inst & $10^6$ & $-1.5$ \\
Pop III.1	      & 1.0 & Inst & $10^7$ & $-1.0$ \\
Pop III.1	      & 0.0 & $10^7$ & $10^7$ & $-1.1$ \\
Pop III.1	      & 1.0 & $10^7$ & $10^7$ & $-0.6$ \\

Pop III.2	      & 0.0 & Inst & $10^7$ & $-1.2$ \\
Pop III.2	      & 0.0 & Inst & $10^6$ & $-1.0$  \\
Pop III.2	      & 1.0 & Inst & $10^7$ & $-0.7$ \\
Pop III.2	      & 0.0 & $10^7$ & $10^7$ & $-0.8$  \\
Pop III.2	      & 1.0 & $10^7$ & $10^7$ & $-0.4$  \\

Pop III, Kroupa & 0.0 & Inst & $10^7$ & $-0.5$ \\
Pop III, Kroupa & 0.0 & Inst & $10^6$ & $-0.5$  \\
Pop III, Kroupa & 1.0 & Inst & $10^7$ & $-0.1$  \\
Pop III, Kroupa & 0.0 & $10^7$ & $10^7$ & $-0.2$ \\
Pop III, Kroupa & 1.0 & $10^7$ & $10^7$ & --  \\
\hline
\label{limits}
\end{tabular}
\end{table}

\section{Discussion}
\label{discussion}
\subsection{Lensed versus unlensed fields}
\label{lensed_vs_unlensed}
\begin{figure}
\includegraphics[width=84mm]{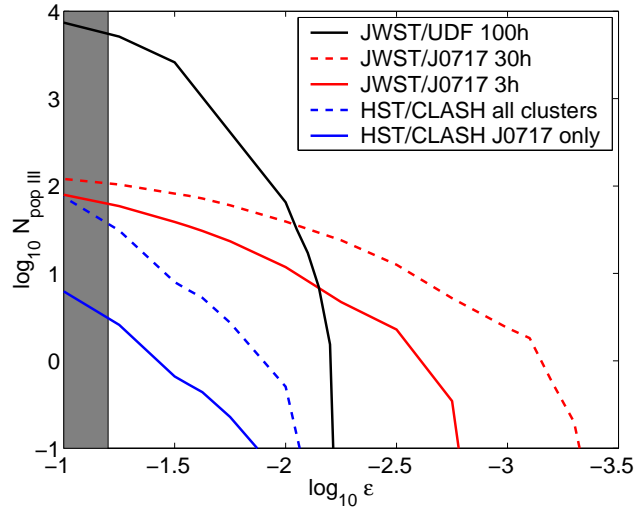}
\caption{The number of pop III galaxy images predicted above the $5\sigma$ detection thresholds in different surveys, as a function of the pop III star formation efficiency $\epsilon$. The blue lines represent the expected results from the {\it HST/CLASH} images of the J0717 galaxy cluster (blue solid) and of all the {\it CLASH} clusters (blue dashed) in the F160W filter. The red lines indicate the corresponding results from {\it JWST} F200W images with 3h (red solid) and 30h (red dashed) exposure times of J0717. These  lensing surveys are contrasted with the corresponding number of pop III galaxies predicted for a {\it JWST/UDF} 100h survey (black line) of an unlensed field. A model with a Pop III.2 IMF, $f_\mathrm{esc}=0$ and $\tau_\mathrm{popIII}=10^7$ yr is assumed. The gray region represents the SFE range ruled out by existing LF constraints (Sect.~\ref{LF}). While the {\it JWST/UDF} images are expected to produce far more pop III galaxies than the J0717 observations at high star formation efficiencies, only J0717 observations can probe pop III galaxies at $\log_{10} \epsilon\lesssim -2.5$.}
\label{CLASH_vs_Palantir}
\end{figure}

In Fig.~\ref{CLASH_vs_Palantir}, we compare the expected performance of both {\it HST/CLASH} and our hypothetical {\it JWST} J0717 survey to that of ultradeep ($t_\mathrm{exp}=100$ h) {\it JWST/NIRCam} observations of an unlensed field, by plotting the total number of pop III galaxy images $N_\mathrm{popIII}$ above the survey $5\sigma$ detection thresholds as a function of the pop III galaxy SFE. The results in this figure are based on a model with a Pop III.2 IMF, $f_\mathrm{esc}=0$ and $\tau_\mathrm{popIII}=10^7$ yr, but the relative difference between the survey results are fairly similar for all models considered. Since lensed fields cover a smaller background volume, more pop III galaxies are predicted in the {\it JWST} UDF at high SFEs, but the numbers drop more rapidly with decreasing SFE than in the {\it JWST} J0717 observations. For this particular pop III galaxy model, only {\it JWST} observations of J0717 are able to reach $\log_{10} \epsilon \lesssim -2.5$ with $N_\mathrm{popIII}\gtrsim 1$ (i.e. $\log_{10} N_\mathrm{popIII}\gtrsim 0$). In relative numbers, a {\it JWST} survey of J0717 probes SFEs 0.4 dex (3 h exposures) and 1.0 dex (30 h exposures) deeper than the {\it JWST} UDF. 

Included in Fig.~\ref{CLASH_vs_Palantir} are also the pop III image counts predicted for the {\it HST/CLASH} survey. A direct comparison between the {\it HST} and {\it JWST} results for J0717 indicates that the {\it JWST} survey of this cluster would be able to probe SFEs (or, equivalently, intrinsic luminosities) that are lower by $\approx 1.2$ dex (3 h exposures) and $\approx 1.8$ dex (30 h exposures). When compared to the expected performance of the full {\it HST/CLASH} sample, this {\it JWST} survey instead reaches $\approx 0.7$ dex (3 h exposures) and $\approx 1.3$ dex (30 h exposures) deeper in terms of SFE. In summary, observations of a high-magnification cluster like J0717 should offer the best chances of detecting pop III galaxies with {\it JWST}.

\subsection{Image multiplicity}
In situations where our lensing simulations predict $\log_{10} N_\mathrm{popIII}(m \leq m_\mathrm{limit})>0$ (i.e. more than one detectable image), the number of lensed pop III galaxy images will typically not equal the number of {\it unique} pop III galaxies detectable in the surveys we consider, since a fraction of the images will stem from the same source. We find that the image multiplicities of our mock samples usually lie in the range $\approx 1.2$--2. Hence, our $N_\mathrm{popIII}$ estimates in Figs.~\ref{Palantirfig}, ~\ref{CLASHfig},  and \ref{CLASH_vs_Palantir} should be divided by a factor in this range to recover the number of unique pop III galaxies. In the case of the unlensed {\it JWST/UDF} field (Fig.~\ref{Palantirfig}b), no correction is of course required. The reason why we present the number of images, rather than the number of unique galaxies, is that having more images tends to be advantageous -- some of the images (both in lensed and unlensed surveys) may be rendered useless due to blending with foreground objects (Milky Way stars and low-redshift galaxies).

In principle, the image multiplicity depends on the positions of the sources with respect to the caustics of the lensing cluster. However, our simulations also indicate a correlation between image multiplicity and the magnifications of the images that make it above the detection threshold. This makes the typical image multiplicities sensitive to the intrinsic luminosities of the pop III galaxies which, in turn, depend on the model parameters described in Sect.~\ref{SEM}. The fact that the image multiplicity in many cases is $<2$ indicates that not all lensed pop III galaxies produce detectable multiple images. There are several reasons for this. To begin with, not all highly magnified pop III galaxies are multiple imaged -- our lens mapping predicts a significant fraction of single images close to but outside the caustics (e.g. $\approx 50$\% of all images with $\mu\gtrsim 20$). Secondly, the flux ratios of the multiple images are sometimes too large to allow for more than one to make it above the detection threshold -- flux ratios of up to a factor of $\approx 2$ (corresponding to $\approx 0.75$ magnitudes) between the brightest and second brightest images are not uncommon. Finally, some multiply imaged systems display such large image separations ($\gtrsim 1\arcmin$) that even though several images are in principle detectable in the J0717 field (Fig.~\ref{J0717}), only a single one appears within the part of the cluster that we assumed to be covered by the observations.

\subsection{Star formation history and duration of the pop III phase}
\label{secondary_parameters}
The results presented throughout Sect.~\ref{Palantir} and ~\ref{CLASH} are based on the assumption that pop III galaxies form their stars more or less instantaneously and regain their pop III spectral signatures for $\tau_\mathrm{popIII}=10^7$ yr.

Contrary, perhaps, to naive expectation, the $\tau_\mathrm{popIII}$ parameter is not necessarily of critical importance for the prospects of detecting pop III galaxies in the surveys we consider. While reducing $\tau_\mathrm{popIII}$ from $10^7$ yr to $10^6$ yr formally reduces the number densities of all pop III galaxies by an order of magnitude\footnote{For pop III.1 galaxies, which have total lifetimes of as little as $\approx 3$--4 Myr, the correction factor is closer to $\approx 1/3$} at any redshift, the numbers of objects that appear above the relevant detection thresholds in figures like Fig.~\ref{Palantirfig} and ~\ref{CLASHfig} typically change by less than $\approx 50\%$. This happens because it is only the very brightest pop III galaxies that have any sporting chance of being detected. Since pop III galaxies experiencing an instantaneous burst of star formation attain their peak luminosities already in $\approx 1$--$2\times 10^6$ yr (depending on exact IMF and filter), removing objects with ages above this limit has little or no impact on the high-luminosity tail of the intrinsic pop III luminosity distribution. This effect can be readily seen by comparing the intrinsic luminosity functions for pop III.2, $f_\mathrm{esc}=0$, $\log_{10} \epsilon = -1.5$ galaxies with $\tau_\mathrm{popIII} = 10^7$ yr (orange line in Fig.~\ref{LFfig_1}) and $\tau_\mathrm{popIII} = 10^6$ yr (green line in Fig.~\ref{LFfig_2}) -- the two distributions are similar at the high-luminosity end but very different at lower luminosities. Gravitational lensing can in principle overthrow this trend by assigning high magnifications to intrinsically faint galaxies and low magnifications to intrinsically bright galaxies, but in our simulations, this mechanism is not sufficiently efficient to make $\tau_\mathrm{popIII}$ a crucial parameter in the current context.

Extending the star formation history has the effect of lowering the peak luminosity per unit star formation efficiency and to prolong the brightest phase in the life of these objects. This makes all simulated pop III galaxies less luminous but more numerous (as long as $\tau_\mathrm{popIII}$ is not shorter than the duration of the starburst). When it comes to the prospects of detecting rare objects in lensed fields, boosting the intrinsic (unlensed) object number densities could in principle be advantageous, but in our simulations it does more harm than good, partly because we impose a $\mu\leq 100$ threshold. When going from an instanstaneous burst to a $10^7$ yr star formation episode, we find that the SFE on average needs to be boosted by $\Delta \log_{10} \epsilon \approx 0.2$--0.4 dex (i.e. a factor of $\approx 2$--3) to produce the same number of objects above the detection thresholds in our lensed {\it HST} and {\it JWST} surveys.

\subsection{Uncertainties in the lensing maps}
\label{uncertainty_section}
The \citet{Zitrin et al. a} J0717 magnification maps used in our simulations come with a certain level of uncertainty, and this translates into an uncertainty in our estimates of the number of detectable pop III galaxies in the {\it HST} and {\it JWST} surveys of the cluster we consider. Since the deflection field scales with the lens-source ($D_\mathrm{ls}$) to source ($D_\mathrm{s}$) angular-size distance ratios,  cluster lensing maps require knowledge of at least one multiply-lensed source redshift to calibrate the mass model, and of other redshifts in order to deduce the mass profile and accurately assess the magnification. However, by the time our J0717 magnification maps were produced, no such spectroscopic redshifts were available. Based on only two available imaging bands, \citet{Zitrin et al. a} instead estimated a redshift of $z\approx 2$--2.5 for the main source to which their model was calibrated (system $\#1$ in their notation), whereas other dropout multiply-lensed objects found in their work were used to pin down the profile and the resulting magnification. The \citet{Zitrin et al. a} lensing maps, and by extension our simulation results, are therefore implicitly based on the assumption that the redshift of system $\#1$ is $z\approx 2$. Recently, the first spectroscopic information was published for objects behind cluster, revealing that $\#1$ in fact is at a redshift of $z=2.963$ \citep{Limousin et al.}. This entails two possible systematic errors which we now make an effort to assess. 

First, the lensing distance ratios for higher-$z$ source planes should now be slightly lower. Luckily, due to the converging nature of the $D_\mathrm{ls}/D_\mathrm{s}$ for higher source redshift for a fixed lens, the corrections for the higher-redshift source plane angular-size distances are minor and decrease with increasing source redshifts. For example, assuming that $\#1$ is at $z=2$ as we did, results in a \emph{relative} $D_\mathrm{ls}/D_\mathrm{s}$ factor of $\approx 1.3$ for a source-plane redshift of $z_\mathrm{s}=10$. By instead using the correct redshift of $z_\mathrm{s}=2.963$, the relative ratio becomes $\approx 1.2$. To assess the impact of this on our results, we have produced magnification maps for a typical redshift of $z_\mathrm{s}=10$ according to both $\#1$ redshift options and compared the magnification properties. We find that the area of $\mu>10$ we used in our calculations is only $8\%$ larger; and for a threshold of $\mu>20$ or higher, larger by $\lesssim 1\%$, therefore rendering a possible systematic bias negligible. 

Second, due to the higher redshifts of system $\#1$ and $\#5$ than was assumed in \citet{Zitrin et al. a}, the cluster mass profile might also be affected at some level. Due to the extended merging nature of this cluster, the profile is still expected to be very shallow, so that any impact on the mass profile is correspondingly expected to be very small. To examine this effect, we have checked that within the given uncertainties, our model is commensurate with the new redshift measures for the systems found in \citet{Zitrin et al. a}, especially systems $\#3$ and $\#5$ which expand the redshift range, and reproduces these systems at their given redshifts. We find a general good agreement, although we note that the mass profile of the model in use may be slightly shallower in the very inner core than needed to perfectly match the $D_\mathrm{ls}/D_\mathrm{s}$ growth. We therefore compare the high magnification areas with a variety of steeper models, and find that the size of the high-magnification areas are subject to a $\lesssim 10\%$ uncertainty because of this. 

Since the expected number of detectable pop III galaxies is proportional to the size of these high-magnification regions, we conclude that the new redshift information provided by \citet{Limousin et al.} would result in very small adjustments of our predictions. A further refinement of the mass model will soon also be enabled by the {\it CLASH} 16-band imaging of J0717 and the resulting photometric redshifts for its multiply-lensed systems.

\subsection{The colour signatures of pop III galaxies}
\label{colours}
Throughout this paper, we have explored the pop III galaxy properties required to place them above the flux detection thresholds of different {\it HST} and {\it JWST} surveys. However, having a galaxy bright enough for detection does not necessarily mean that one will also be able to identify it as a pop III object based on its colours. In fact, only certain subtypes of pop III galaxies are likely to display colour signatures that set them apart from from more mundane objects.   

Like all galaxies in the reionization epoch ($z\gtrsim 6$), pop III objects will appear in dropout-selected samples because of absorption shortward of the Ly$\alpha$ line by the largely neutral IGM. However, additional colour criteria are also required, and these depend on the exact properties of the pop III galaxies. Pop III galaxies dominated by direct star light can be uniquely identified for ages at least up to $\approx 10$ Myr in {\it HST} or {\it JWST} filters because of their very blue UV-slopes \citep[e.g.][]{Schaerer & Pello,Raiter et al.,Inoue,Zackrisson et al. b}. Pop III galaxies dominated by nebular emission (all our $f_\mathrm{esc}=0$, $\tau_\mathrm{popIII}\leq 10^7$ yr models) can be identified from their {\it JWST} colours because of their very strong hydrogen Balmer emission lines but lack of [OIII] emission \citep{Inoue,Zackrisson et al. b}. Unfortunately, the colour signatures arising from this are unique only over very limited redshift intervals. The [OIII] line at 5007 \AA{} also redshifts out of the wavelength range of the {\it JWST/NIRCam} instrument at $z\gtrsim 8$, and even though the {\it JWST/MIRI} camera can be used to probe this line at higher redshift, the lower sensitivity of this instrument makes this endeavour extremely expensive in terms of exposure time \citep{Zackrisson et al. b}. In general, we find that {\it MIRI} detections of objects appearing at the {\it NIRCam} detection limit (Fig.~\ref{Palantirfig}) would require such long exposure times ($\gtrsim 100$ h per filter) that this approach seems unrealistic. Pop III galaxies with nebular emission can also exhibit extremely strong Ly$\alpha$ emission \citep[]{Schaerer a,Schaerer b,Raiter et al.}, and this can in principle produce distinct colour signatures in either {\it HST} (\citealt{Zackrisson et al. c}) or {\it JWST} filters. However, since Ly$\alpha$ is readily absorbed in the increasingly neutral IGM at $z\gtrsim 6$, this requires a fortunate sightline and can most likely apply only to the objects at the lowest redshifts considered in our simulations. 

Realistically, colour criteria can probably only be used to select pop III galaxy {\it candidates}, whereas follow-up spectroscopy will be required to achieve a more robust identification of pop III objects, based either on UV continuum slope or the strength of emission lines like Ly$\alpha$, HeII, H$\beta$, H$\alpha$ and [OIII] \citep{Schaerer a,Schaerer b,Schaerer & Pello,Raiter et al.,Inoue,Cai et al.,Zackrisson et al. c}. Fortunately, objects appearing above the detection thresholds of {\it HST/CLASH} ($m_\mathrm{AB}\lesssim 27$) are within range of existing near-IR spectrometers on 8--10 m groundbased telescopes \citep[e.g.][]{Schenker et al.,Lehnert et al.}, and objects appearing above the detection limits of a lensed {\it JWST} survey ($m_\mathrm{AB}\lesssim 29$ for $t_\mathrm{exp}=3$ h) are within the range of the {\it JWST/NIRSpec} spectrometer or planned groundbased telescopes like the {\it Giant Magellan Telescope}\footnote{http://www.gmto.org/}, the {\it Thirty Meter Telescope}\footnote{http://www.tmt.org/} and the {\it European Extremely Large Telescope}\footnote{http://www.eso.org/sci/facilities/eelt/}. The same exercise would be far more difficult for pop III galaxy candidates with comparable SFEs detected in an unlensed field like the {\it JWST/UDF}. For example, the lowest SFE at which the {\it JWST/UDF} still produces competitive pop III galaxy counts compared to a lensed {\it JWST} survey (see Fig.~\ref{CLASH_vs_Palantir}) converts into an unlensed pop III galaxy flux of $m_\mathrm{AB}\approx 30.5$--31.0, which would be very challenging to probe spectroscopically in the foreseeable future. 

\subsection{Hybrid pop III galaxies}
Our simulations trace the formation of metal-free halos that are likely to host {\it pure} pop III galaxies, but due to incomplete mixing of metals in the interstellar medium, pop III star formation may plausibly continue at some minor level within objects that have already started to form pop I/II stars \citep[e.g.][]{Johnson et al. 09,Salvaterra et al.}. In most cases, this will dilute any pop III spectral or colour signatures beyond recognition, but certain diagnostic criteria, like the strength of the Ly$\alpha$ line, can in principle apply even if just a minor fraction of the overall stellar mass of a high-redshift galaxy is in the form of pop III stars \citep{Zackrisson et al. c}. Such hybrid pop III galaxies are not included in our source count estimates, and could boost the prospects of detecting pop III galaxies in both lensed and unlensed fields.   

\section{Summary}
Using simulations of the fluxes of pop III galaxies, we argue that the observed galaxy luminosity function at $z=7-10$ can be used to place upper limits on the typical star formation efficiencies of these objects, and present such limits for a wide range of model parameters. By projecting these simulations through the magnification map of the galaxy cluster J0717 (arguably the best lensing cluster currently known for high-redshift surveys), we predict that the ongoing {\it HST} imaging survey {\it CLASH}, which targets a total of 25 galaxy clusters -- including the J0717  -- potentially could turn up a small number of pop III galaxies if the typical star formation efficiency is $\sim 10^{-2}$.  We also explore the prospects of surveying the J0717 cluster with the upcoming {\it JWST} telescope, and find that this strategy would be able to detect pop III galaxies with star formation efficiencies an order of magnitude lower ($\sim 10^{-3}$ of the baryons converted into pop III stars). The gravitationally enhanced fluxes of such objects also make them potentially suitable targets for follow-up spectroscopy, whereas this would be exceedingly difficult for corresponding objects in unlensed fields.

\section{acknowledgements}E.Z, C-E.R and G.\"O. acknowledge funding from the Swedish National Space Board. E.Z. and G.\"O. also acknowledge funding from the Swedish Research Council. A.Z. is supported by the ``Internationale Spitzenforschung II/2'' of the Baden-W\"urttemberg Stiftung. C-E.R acknowledges funding from the Royal Swedish Academy of Sciences. The anonymous referee is acknowledged for insightful comments which helped improve the quality of the paper.

\end{document}